\documentclass[aps,floatfix,showpacs,10pt]{revtex4}
\usepackage[dvips]{graphicx}
\newcommand{\gst}{\ensuremath{\mathrm{Ge_{2}Sb_{2}Te_{5}}} }
\newcommand{\gstb}{\ensuremath{\mathbf{Ge_{2}Sb_{2}Te_{5}}} }

\newcommand{\mmr}[1]{\ensuremath{\mathrm{#1}} }

\begin{document}

\title[] {A possible mechanism of ultrafast amorphization in phase-change 
memory alloys:
an ion slingshot from the crystalline to amorphous position.
\footnote{Published: J.\ Phys.: Condens. Matter \textbf{19} (2007) 
455209 (7pp). \copyright 2007 IOP Publishing LTD.}}

\author{A. V. Kolobov$^{1,2}$, A. S. Mishchenko$^{3,4}$, P. Fons$^{1}$, S.M. 
Yakubenya$^{4}$, J. Tominaga$^{1}$}


\affiliation{$^1$Center for Applied Near-Field Optics Research (CanFor), 
National Institute of Advanced Industrial Science and Technology, 
1-1-1, Higashi, Tsukuba 305-8562, Japan
\\ 
$^2$Institut Charles Gerhardt, UMR 5253 CNRS-UM2-ENSCM-UM1, PMDP, 
Universit\'{e} Montpellier II, 
Place Eug\`{e}ne Bataillon, Montlpellier Cedex 5, France
\\
$^3$CREST, Japan Science and Technology Agency (JST), 
AIST, 1-1-1, Higashi, Tsukuba 305-8562, Japan
\\
$^4$ RRC ``Kurchatov Institute", 123182, Moscow, Russia 
}

\begin{abstract}
We propose that the driving force of an ultrafast crystalline-to-amorphous 
transition in phase-change memory alloys are strained bonds existing in the 
(metastable) crystalline phase. For the prototypical example of \gst, we 
demonstrate that upon breaking of long Ge-Te bond by photoexcitation  
Ge ion shot from an octahedral crystalline to a tetrahedral amorphous 
position by the uncompensated force of strained short bonds.   Subsequent 
lattice relaxation stabilizes the tetrahedral surroundings of the Ge atoms 
and ensures the long-term stability of the optically induced phase.
 \end{abstract}
%

\maketitle

\section{Introduction}
Photo-induced phase transitions have attracted ongoing attention because of 
their application in storage devices \cite{NazuBook,KolBook}. Properties 
required for commercial memory media, 
such as fast switching, stability of the photo-converted state, and high 
level cyclability singled out phase-change effect in Te-based chalcogenides.  
Commercially utilized materials are Ge-Sb-Te alloys mainly in the form of 
\gst (GST) used in digital versatile discs (DVDs) DVD-RAM and Ag-In-Sb-Te 
(AIST) alloys used in DVD-RW \cite{Ohta}.  \gst stands over a million cycles, 
the switching time is on the nanosecond time scale and even femtosecond 
pulses are sufficient to render the material amorphous \cite{Ohta}. 
    
As is typical for many optically sensitive materials 
\cite{NazuBook,KolBook,TokuraPRL}, amorphous chalcogenides are intrinsically 
metastable and undergo various structural transformations under the influence 
of an external perturbation, in particular light. What makes these materials 
special? It is very difficult to believe that a material will stand millions 
of melting/solidification cycles without degradation of parameters and an 
essentially specific mechanism should be found.  The fundamentals of phase-
change mechanism are just starting to emerge but striking features of both 
groups of materials are: (i) the existence of longer and shorter bonds 
between similar pairs of atom types in the crystalline state \cite{A7,NatMat}, 
(ii) bond shortening in the amorphous state \cite{A7,NatMat,Sb2Te3}, and 
(iii) a considerable degree of intrinsic disorder, manifested in high 
concentrations of vacancies in GST \cite {Yamada} and a random occupancy of 
sites in AIST \cite{A7}. The first feature suggests that certain bonds in the 
system are weaker and can be selectively broken, the second one is an 
indication of the local atomic structure in the two states being 
significantly different, and the last property points to considerable 
concentration of defect levels in the gap.

In this Letter, we propose a model that explains the unusual features of the 
structural changes in the class of phase-transition chalcogenides. We 
concentrate on changes in a typical compound, \gst, although our generic 
statements, based on the similarity of  properties (i)-(iii), are general for 
all phase-
change compounds.

\section{The structure of \gstb}

In the metastable form of GST Te ions form a well ordered fcc sublattice, 
with Ge and Sb being displaced from the center of the cell 
\cite{NatMat,Sha,Wutt}. As a result, there are subsystems of shorter and 
longer bonds. The shorter Ge(Sb)-Te bond lengths are 2.83~\AA~ and 2.91~\AA , 
respectively. The longer bonds are on the order of $\sim 3.15$~\AA. In 
addition, there are 20 percent vacancies on the Ge/Sb sites that are 
intrinsic to the structure \cite{NatMat} but the role and exact location of 
which is not definitely known. It should be stressed, that the intermediate- 
and long-range order of the crystal structure are not finally clear the best 
description to-date being distorted rock-salt like structure with large 
isotropic thermal displacement parameter.

\begin{figure}[bht]
	\centering
\includegraphics[width=8 cm]{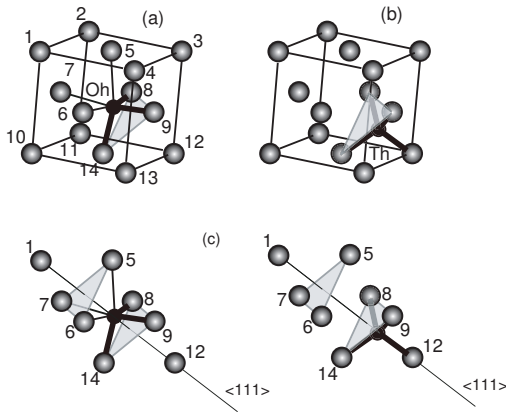}
\caption{\label{fig:fig1} Fragments of the GST structures with the Te fcc 
lattice and a Ge atom located in the octahedral (a) and tetrahedral (b) 
symmetry sites. Other Ge (and Sb) atoms are not shown. The lower panel (c) 
shows the molecular fragments (8 Te atoms and 1 Ge atom) within which the Ge 
switching is considered.
}
\end{figure}

\subsection{Photoexcitation}
It was shown \cite{NatMat} that upon exposure to intense laser pulses  Ge 
atoms switch from an octahedral position at the center of the rocksalt cell 
into a tetrahedral position within the Te fcc sublattice (Fig. 1a,b) which is 
responsible for the fast performance and high stability of the medium. It was 
suggested \cite{NatMat}  and subsequently unequivocally demonstrated by {\it 
ab initio} simulations \cite{Wutt} that the total energies of Ge-Sb-Te for 
these two structures are rather close which explains why both states possess 
high stability. However, a very important question remained unanswered, 
namely, what drives the Ge atom to switch to a tetrahedral symmetry position 
upon photoexcitation? 

It should be noted that even the short bonds in GST (2.83~\AA~   for Ge-Te 
and 2.91~\AA~  for Sb-Te) are significantly longer that the sum of the 
corresponding covalent radii ($r_{\mbox{\scriptsize Ge}} = 1.22$~\AA, 
$r_{\mbox{\scriptsize Sb}} = 1.38$~\AA, $r_{\mbox{\scriptsize Te}} = 
1.35$~\AA~ \cite{webelements}), i.e. all bonds - assuming they are covalent - 
are stretched. This is a crucial assumption that has to be justified. To 
address this issue, we have measured bulk XPS spectra of \gst using high-
energy synchrotron radiation as the excitation source. Our measurements 
yielded the following values for the bonding energies: Ge 3d - 30.8 eV, Sb 
3d$_{5/2}$ - 529.6 eV, Te 3d$_{5/2}$ - 573.2 eV. These values are similar to 
those in covalently bonded solids \cite{XPSHandbook}. The absence of charge 
transfer has been independently deduced from {\it ab initio} simulations 
\cite{Wutt, Robertson1}. 

It would seem that a simple decrease in the lattice parameter would have 
reduced the strains thus lowering the total energy. However, the system 
prefers a larger lattice parameter and stretched (strained) bonds. The most 
plausible explanation for this is the fact that the Te fcc sublattice is 
intrinsically much stronger than the Ge(Sb)-Te interaction \cite{NatMat}.  In 
line with this is also the fact that Si-Te bond lengths for octahedrally and 
tetrahedrally coordinated Si in Si$_2$Te$_3$ have the same lengths 
\cite{SiSbTe} as those for Ge-Te, although Si has a  covalent radius that is 
significantly (0.11~\AA ) smaller than that of Ge. 

\subsection{A simple energy calculation}

As already stated above, the details of the structure beyond the short-range 
order (i.e. including just a few atoms) are not known in either crystalline 
or amorphous cases. For this reason any reliable modern solid-state 
calculations are not possible: one is naturally limited to a very crude 
short-range order model. To the first approximation we can consider that Ge 
ions switch within a rigid Te sublattice as shown in Fig. 1a,b. As the 
simplest model we can consider motion of a relatively light Ge ion within the 
eight nearest (and heavy) Te atoms (Fig. 1c). 

Based on very simple considerations the potential experienced by the Ge ion 
located at distance $d$ from a Te ion can be expressed in terms of a standard 
ion-ion interaction relation:
\begin{equation}
E(d) = C/d^4 \pm 2 \sqrt{ V_2^2 + V_3^2} \; ,
\label{hari}
\end{equation}  
where $V_2 = A/d^2$ is the covalent energy, $V_3$ is the independent of 
distance polar energy, and the first term is the interionic repulsion 
\cite{Harbook,HarCi}. In equation (\ref{hari}) the plus and minus signs 
correspond to bonding and antibonding states, respectively.

To determine the parameters of the interaction we used covalent radii 
$r_{\mbox{\scriptsize Ge}}=1.22$~\AA~and  $r_{\mbox{\scriptsize 
Te}}=1.35$~\AA~to get the equilibrium of potential at $d_0=2.57$~\AA~ which, 
implying the condition $\partial E(d)/ \partial d = 0$, fixes the repulsion 
parameter$C=A^2 / \sqrt{V_3^2 + A^2/d_0^4}$. The polar energy $V_3=1.66$ eV 
is determined in Ref.~\cite{HarCi} and a reasonable value of covalent 
attraction $V_2=2$ eV \cite{HarCi} in equilibrium gives A=13.2~eV~\AA$^2$. 
With the above parameters we have calculated the potential for a Ge atom 
moving within a rigid Te lattice along the $[111]$ direction as shown in Fig. 
1c.

\begin{figure}[bht]
	\centering
\includegraphics[width=8 cm]{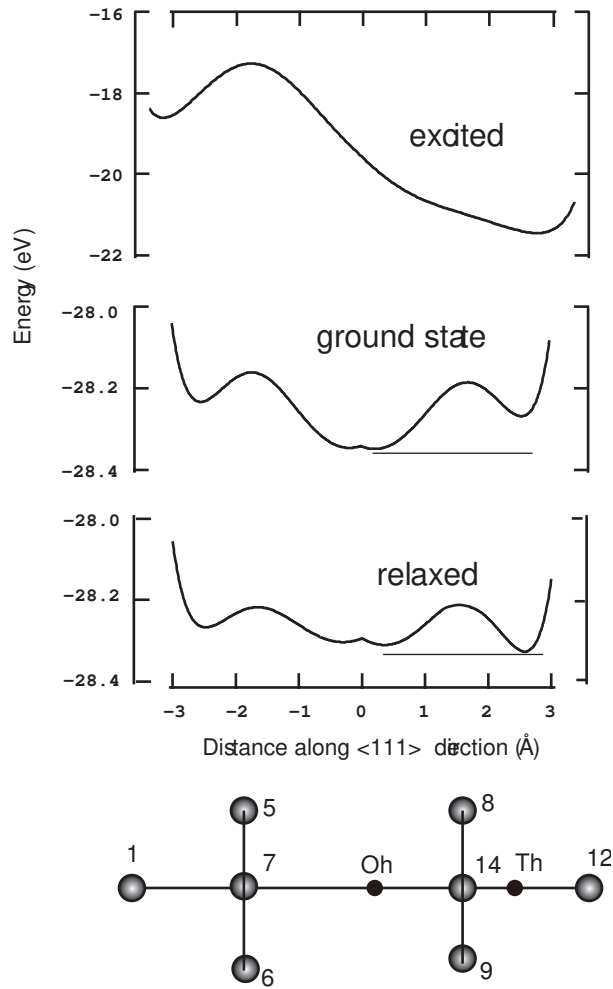}
\caption{\label{fig:fig2} Potentials for the Ge atom in the field 
of eight Te atoms in the ground state and in the state where 
one of the weak bonds is excited (upper panel) \cite{note3}. The schematic 
drawing
represents the considered molecule. The lower panel demonstrates the 
stabilisation of the tetrahedral geometry by relaxation of the network (Te 
atoms 
5,6,7 have been moved away in the direction perpendicular to the axis by 
0.1~\AA . 
}
\end{figure}

The 'ground state' curve in Fig. 2 presents the potential in the ground state. 
The origin of the coordinate $r$  - which represents the position of the Ge 
atom along the 111 axis - corresponds to the centrosymmetric position of Ge 
between the two Te triangles shown in Fig. 1c. The experimental positions of 
Ge are $r_{\mbox{\scriptsize cr}}=0.33$~\AA~in the crystalline state and 
$r_{\mbox{\scriptsize am}}=2.6$~\AA~in the amorphous state.

It is seen that the very simple form of interaction (\ref{hari}) reproduces 
general features of the system with the crystalline global minimum  at $r 
\approx 0$ and higher amorphous local minima at $r \approx 2.6$~\AA. Besides, 
although the minima of potential (\ref{hari}) from three Te ions (located off 
the [111] axis) to the right (left) correspond to r=0.86~\AA~ (r=-0.86~\AA), 
the joint action of left and right Te ions creates a broad minimum at $r 
\approx 0$.    

From the Jahn-Teller theorem it is known \cite{Bersuker,Kristoffel}that 
centrosymmetric structures are locally unstable and centrosymmetric positions 
in solids are allowed as a consequence of long-range stabilizing forces that 
occur within a macroscopic periodic lattice.  
For the case of GST, the high concentration (20\%)  of vacancies destroys the 
long-range order leading to the formation of the non-centrosymmetric 
distorted rocksalt structure.    This is evidenced by the loss of long-range 
order in the Ge/Sb sublattice, as opposed to the well defined Te sublattice 
where long-range spatial correlations are clearly visible  \cite{NatMat}.

To reproduce the displacement of the Ge atom from the centrosymmetric 
position we introduce Jahn-Teller coupling \mmr{\lambda} and a coupling to an 
external electric field $\xi$ caused by non-symetric distribution of 
vacancies in the Ge/Sb sublattice. The above interactions create an 
additional potential \cite{Bersuker}
\begin{equation}
\Delta E(r) = - \sqrt{{\cal N}^2 + \lambda^2 r^2} - \xi r \; .
\label{dobavki}
\end{equation} 
We chose the values \mmr{\lambda=0.08} eV~\AA$^{-1}$ and \mmr{\xi = 
0.007}~eV\AA$^{-1}$ to reproduce the experimental position of Ge. The non-
adiabatic matrix element ${\cal N} = 0.0006$ eV does 
not change the potential significantly though gives possibility to avoid an 
unphysical jump of the second derivative of potential at r=0. 

\begin{figure}[h]
	\centering
\includegraphics[width=8 cm]{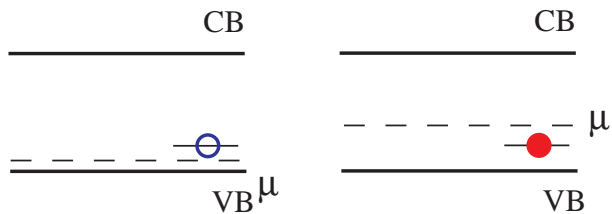}
	\caption{Schematic energy band structure of \gst at low (left) and high 
(right) 
excitation levels.
}
	\label{fig:bandstructure}
\end{figure}

When an electron on one of the (long) bonds is transferred to the antibonding 
orbital due to photoexcitation \cite{note2}, the potential changes 
drastically (Fig. 2, the 'excited' curve). In particular, in the excited 
state there appears a force acting on the Ge ion. The excited Ge ion relaxes 
into a position corresponding to the minimum located at 2.6~\AA , i.e. at the 
tetrahedral position.  We estimated the relaxation time to be on the scale of 
$10^{-13}$ s (the inverse of the phonon frequency) and the atom travels a 
distance of about 2~\AA . A simple estimate gives an impressive speed of 
$\approx 10^3$ m/s. A simple analogy of the process is a slingshot shooting a 
Ge atom.

At the same time, the (5,6,7) Te atoms that were previously bonded to the Ge 
atom by the longer bonds, experience a stronger interaction with the Sb atoms 
located on the other side (Sb atoms are not shown in the Figure). The Sb-Te 
bonds shrink by about 0.1~\AA~as evidenced by the EXAFS data \cite{NatMat},  
resulting in a Te lattice distortion (See Fig. 5 of ref. \cite{KolJJAP}). If 
the three (5,6,7) Te atoms are moved away from the axis by 0.1~\AA , the 
tetrahedral geometry becomes the lowest in energy as shown in the lower panel 
of Fig. 2 ('relaxed' curve). The lattice relaxation thus stabilises the 
tetrahedral position of the Ge atom. The energy barrier, separating the 
octahedral and tetrahedral sites is on the order of 0.1-0.2 eV in our simple 
model which is similar to the value obtained through advanced {\it ab-initio} 
simulations  \cite{Robertson2}. 

\subsection{Laser excitation}
An important issue is why one needs intense laser pulses to induce the phase 
change. Vacancies in semiconductors are known to produce deep donor states 
\cite{Newton,Baraff}. The presence of a very large concentration of vacancies 
in the \gst structure thus leads to the chemical potential being pinned near 
the valence band.  The local switching  of Ge from octahedral to tetrahedral 
coordination in the dilute 
limit gives rise to complex defects (a vacancy plus an interstitial Ge atom) 
within the rocksalt structure with corresponding states in the gap.  Since 
the total energies of \gst containing Ge atoms in octahedral and tetrahedral 
symmetry positions are very close \cite{Wutt}, we can expect energy levels 
corresponding to individual tetrahedrally coordinated Ge atoms (and 
corresponding vacancies) to be located closer to the valence band. 

At room temperature and at low excitation levels, the states of the complex 
defects are located above the chemical potential and are hence unoccupied as 
seen in Fig.~\ref{fig:bandstructure}, left. In this situation, a single 
excitation event results in an unstable tetrahedral Ge site which readily 
decays back to the original structure. Under intense optical excitation,  the 
chemical potential moves towards mid-gap.  In this situation the defect 
states are located below the chemical potential and are now occupied (stable) 
(Fig. 3, right). Once the concentration of these defects becomes sufficiently 
large and their wavefunctions overlap, the electronic structure changes to 
that of the ``amorphous'' state. 

The same result can - in principle - be achieved by heating. Indeed, our 
recent XANES measurements (not shown here) have demonstrated that upon 
melting Ge atoms acquire tetrahedral symmetry positions. There is, however, a 
very important difference. A focused laser beam excites a very small volume 
and upon cessation of the laser pulse the heat can dissipate fast enough to 
leave the newly created structure stable. With thermal heating, the heated 
volume is large and cannot be quenched fast enough;  the system can overcome 
the energy barrier between the amorphous and crystalline states eventually 
producing the crystalline phase.

We would like to stress here that the transition into the tetrahedral 
structure is localized, i.e. upon rupture of the long Ge-Te bonds {\it 
individual} Ge {\it atoms} switch due to a force acting on them in the 
excited state. This process is extremely fast. To restore the original 
structure, on the other hand, long-range interactions are important. This 
process requires motion of {\it many atoms} and for this reason 
crystallization is a slower process than amorphization. It is possible that 
collectiveness of the crystallization process is the reason why the 
experimentally measured activation energy for the crystallization is higher 
that the higher than the energy barrier for an individual transition.

\section{Conclusion}
To conclude, we suggest that photoexcitation of a weak Ge-Te bond in a 
distorted rocksalt structure produces a net force acting on the Ge ions along 
the $[111]$ direction which drives the phase transition towards the so-called 
amorphous state. The subsequent distortion of the Te lattice stabilizes the 
tetrahedral structure which ensures the stability of the recorded bit. Whilst 
the detailed discussion above refers to a specific material, the similarity 
of crucial for our mechanism properties (i)-(iii) in all phase-transition 
chalcogenides is indicative of a similar transformation mechanism in all of 
them. Namely, the transition is initiated by photoexcitation of the strained 
bonds and subsequent lattice relaxation stabilises the newly established 
local structure. A necessity for shift of the chemical potential towards mid-
gap that is needed to stabilize the tetrahedral sites is the reason why this 
process requires high photon fluxes.

While for the most part we we concentrated on photoexcitation of the weak 
bonds, we believe that the above considerations are also valid for electronic 
memory devices when the material is switched between the two states by 
intense current pulses. Application of high electric fields leads to 
injection of charge carriers from the electrodes that are likely to behave in 
a similar way as photo-excied carriers.

ASM acknowledges support of RFBR grant 04-02-17363a.

\section*{References}

\bibliographystyle{unsrt}

\begin{thebibliography}{99}
\bibitem{note3} Since we have used a very simple cell consisting of just nine 
atoms (1 germanium and 8 tellurium atoms) the obtained energies have 
typically molecular values. As we have used the ground-state parameters, the 
absolute energy values obtained for the excited state are likely to be 
different. In a solid, the energy levels will split into bands but this must 
not effect the 
main characteristics, in particular, the force acting on an excited Ge atom 
is 
much larger that the characteristic energies of the ground-state potential.
\end{thebibliography}

\end{document}